\documentclass[journal=jacsat,manuscript=article]{achemso}
\mciteErrorOnUnknownfalse

\SectionsOn
\SectionNumbersOn
\usepackage{chemformula} %
\usepackage[T1]{fontenc} %
\usepackage[utf8]{inputenc}
\usepackage{textcomp}
\usepackage{booktabs}
\usepackage{xr}
\usepackage{multirow}
\usepackage{comment}
\usepackage[ruled,linesnumbered]{algorithm2e}
\usepackage{amsmath}
\usepackage{amssymb}
\usepackage{mathtools}
\usepackage[graphicx]{realboxes}
\title{}

\DeclareMathOperator*{\argmax}{arg\,max}
\DeclareUnicodeCharacter{2005}{}
\usepackage[utf8]{inputenc}
\DeclareUnicodeCharacter{03BC}{$\mu$}
\DeclareUnicodeCharacter{03C3}{$\sigma$}
\DeclareUnicodeCharacter{03B2}{$\beta$}
\DeclareUnicodeCharacter{03B1}{$\alpha$}
\DeclareUnicodeCharacter{2009}{ }
\DeclareUnicodeCharacter{2212}{-}
\usepackage{xr}
\usepackage{booktabs}
\usepackage{multirow}
\usepackage{comment}

\usepackage{xr}
\makeatletter

\newcommand{\multrow}[2]{ \multirow{2}{*}{\parbox{{#1}cm}{\linespread{1.2}\selectfont {#2}}} } 

\newcommand{\er}[1]{expected cumulative reward}
\newcommand{\rr}{r}
\newcommand{\cc}{c}
\newcommand{\rfor}{u}
\newcommand{\cfor}{v}

\newcommand{\rxnpicked}[1]{\rr_{#1}}
\newcommand{\cpdpicked}[1]{\cc_{#1}}
\newcommand{\rxnfor}[2]{\rfor_{{#1},{#2}}}
\newcommand{\cpdfor}[2]{\cfor_{{#1},{#2}}}
\newcommand{\cpdind}{j}
\newcommand{\rxnind}{i}
\newcommand{\tarind}{t}
\newcommand{\smcost}{D}
\newcommand{\rpenalty}{P}
\newcommand{\sms}{\mathcal{S}}
\newcommand{\allrxns}{\mathcal{R}}
\newcommand{\targets}{\mathcal{T}}
\newcommand{\allcpds}{\mathcal{C}}
\newcommand{\reward}{U}
\newcommand{\parents}{\mathcal{P}}
\newcommand{\cycle}{\mathcal{Y}}
\newcommand{\clusind}{k}
\newcommand{\clusters}{\mathcal{A}}
\newcommand{\clusterpicked}{a}
\newcommand{\compoundsincluster}[1]{\mathcal{C}_{#1}}
\newcommand{\maxrxns}{R_{max}}
\newcommand{\budget}{D_{max}}
\newcommand{\maxtars}{T_{max}}
\newcommand{\maxclasses}{M_{max}}
\newcommand{\prunedistance}{N}
\newcommand{\likelihoodscore}{L}
\newcommand{\classind}{l}
\newcommand{\classpicked}{m}
\newcommand{\rxnclasses}{\mathcal{M}}

\usepackage{xr}
\makeatletter
\newcommand*{\addFileDependency}[1]{%
\typeout{(#1)}%
\@addtofilelist{#1}
\IfFileExists{#1}{}{\typeout{No file #1.}}
}\makeatother

\author{Jenna C. Fromer}
\affiliation[Unknown University]
{Department of Chemical Engineering, MIT, Cambridge, MA 02139}
\author{Alexandra D. Volkova}
\affiliation[Unknown University]
{Department of Electrical Engineering and Computer Science, MIT, Cambridge, MA
02139}

\author{Connor W. Coley}
\affiliation[Unknown University]
{Department of Chemical Engineering, MIT, Cambridge, MA 02139}
\alsoaffiliation[Second University]
{Department of Electrical Engineering and Computer Science, MIT, Cambridge, MA
02139}

\email{ccoley@mit.edu}
\title[]{Optimal compound downselection to promote diversity and parallel chemistry}
\abbreviations{}

\begin{document}

\begin{abstract}
Early stage drug discovery and molecular design %
projects often follow %
iterative design-make-test cycles. The selection of which compounds to synthesize from all possible candidate compounds is a complex decision inherent to these design cycles that must weigh multiple factors. %
We build upon the algorithmic downselection framework SPARROW that considers synthetic cost, synthetic feasibility, and compound utility, extending it to address additional critical factors related to the risk of synthesis failure, molecular diversity, and parallel chemistry capabilities. These design considerations further align algorithmic compound selection with the true complexity of this decision-making process, allowing SPARROW to capture a broader set of principles typically reliant on expert chemist intuition. %
The application of these formulations to an exemplary case study highlights SPARROW's ability to promote the selection of diverse batches of compounds whose syntheses are amenable to parallel chemistry. %
\end{abstract} 

\clearpage
\section{Introduction} %

Many advancements in medicine, material science, and agriculture rely on the design of functional small molecules. The discovery of such compounds often involves an iterative cycle of synthesizing and testing new designs. Because many more compounds can be designed than can be synthesized, a key decision that drives molecular design cycles is the selection of which designs to pursue from all possible candidates. 

The selection of which compounds to synthesize or purchase in a molecular design cycle requires the simultaneous consideration of multiple criteria including compound utility, synthetic cost, and molecular diversity. Many computer-aided design tools have emerged to support various aspects of compound design and prioritization. Molecular property prediction models can guide molecular generation \cite{sanchez-lengeling_inverse_2018, meyers_novo_2021, anstine_generative_2023} and support the selection of promising designs from a virtual chemical library \cite{shoichet_virtual_2004, walters_applications_2021}. Retrosynthesis models provide potential synthetic paths to compounds of interest \cite{tu_predictive_2023}, %
while synthetic accessibility or complexity scores offer a more coarse-grained indication of synthesizability \cite{ertl_estimation_2009, coley_scscore_2018, thakkar_retrosynthetic_2021}. Additionally, reaction or synthetic route assessment tools exist to predict synthetic cost, difficulty, and/or feasibility \cite{badowski_selection_2019, kuznetsov_extractionscore_2021, seifrid_routescore_2022, pasquini_linchemin_2023}, which can aid in the selection of an optimal route from a set of options. 
Importantly, these methods have all been developed to assess compounds individually, forgoing any consideration of batch efficiency on cost, for example, using %
divergent synthesis and parallel chemistry. %

We previously developed the Synthesis Planning and Rewards-based Route Optimization Workflow (SPARROW) \cite{fromer_algorithmic_2024} as an algorithmic solution to compound prioritization. SPARROW aims to select a batch of compounds and corresponding synthetic routes using the same motivating factors considered by expert chemists: at a high level, selecting the most promising set of compounds while minimizing synthetic cost. As discussed in the original work \cite{fromer_algorithmic_2024}, SPARROW assumes that the utility of testing specific compounds can be quantified as scalar ``reward'' values with multi-parameter scores, acquisition functions, information gain metrics, or other scoring functions. This framework enables the selection of compound designs and synthetic routes that exploit divergent syntheses and shared intermediates while also preferring fewer, higher-confidence reaction steps. %
Contemporaneous work proposed by \citet{briem_diversity-oriented_2024} is also designed to support compound downselection but does not consider the risk of synthetic failure or the utility associated with specific candidates. 

In this work we introduce optimization approaches to address limitations of existing methods for downselection, including the prior version of SPARROW. %
First, we introduce an expectation value---\er{}---that combines the utility of specific compounds with the risk of their syntheses. Next, we make further modifications to the optimization objective function to promote the selection of a diverse batch of compounds and select routes that are amenable to parallel synthesis. 
We illustrate how these design principles and selection criteria improve SPARROW's alignment with the true complexity of decision-making in molecular design cycles.  Finally, we demonstrate how these optimization approaches may be combined to design diverse chemical libraries that can be synthesized in parallel. 

\section{Results and Discussion}

\begin{figure}
    \centering
    \includegraphics{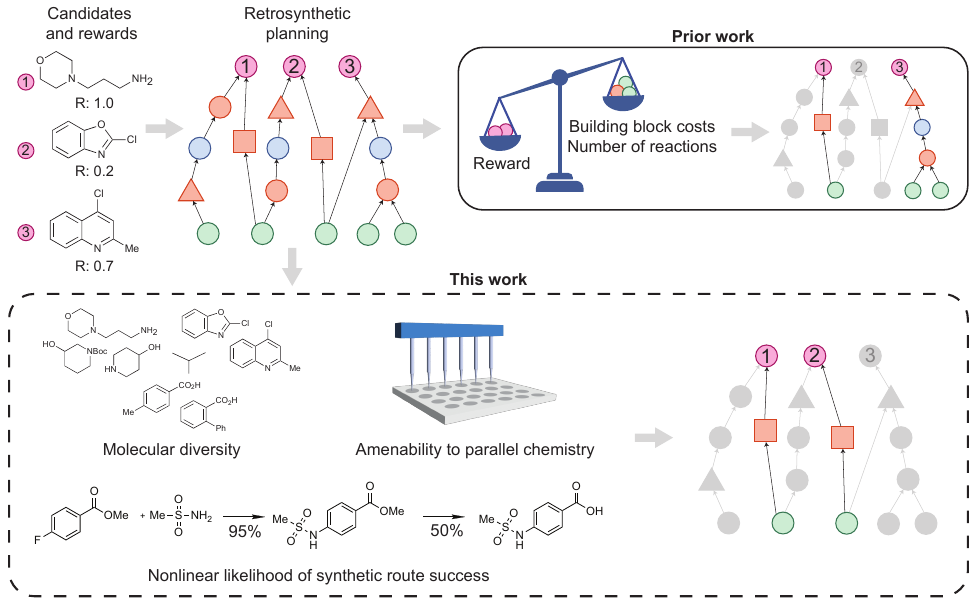}
    \caption{Overview of complexities incorporated into algorithmic compound downselection. We previously developed a framework, SPARROW \cite{fromer_algorithmic_2024}, that downselects batches of compounds and synthetic routes by combining retrosynthesis, molecular property prediction, and prediction of reaction feasibility. In this work, we introduce design principles to promote diversity and amenability to parallel chemistry. We also introduce an expectation value objective, \er{}, that captures the risk associated with synthetic route success in a more principled manner. %
    } %
    \label{fig:overview}
\end{figure}

The capabilities introduced in this work for compound and synthetic route downselection are summarized in Figure \ref{fig:overview}. 
Each of these advancements is incorporated into the open-source algorithmic downselection tool, SPARROW. Details can be found in the Supporting Information.

We demonstrate these improvements on a previously-reported set of 9,621 candidate antibiotics for \textit{Staphylococcus aureus} generated by AutoMolDesigner \cite{shen_automoldesigner_2024}. AutoMolDesigner uses a SMILES-based recursive neural network to generate compounds that optimize predictions of antibacterial activity and other properties relevant to drug design. 
SPARROW uses ASKCOS \cite{tu_askcos_2025} to propose retrosynthetic routes to candidate compounds, predict suitable conditions for candidate reactions, and evaluate candidate reactions in terms of their likelihood of success. With default search settings, ASKCOS identified routes to 3,517 of 9,621 candidates. The reward associated with each candidate compound was defined as the antibacterial score predicted by AutoMolDesigner's binary classifier, bounded between 0 and 1 \cite{shen_automoldesigner_2024}. 
Additional details are provided in Section \ref{sec:methods_network_construction}. 

\subsection{Integrating compound rewards and risk of synthetic failure into a nonlinear expected cumulative reward}

SPARROW's original formulation optimized a linear weighted-sum objective that maximizes the sum of compound rewards, minimizes starting material cost, and minimizes penalties associated with low-confidence reactions\cite{fromer_algorithmic_2024} (Section \ref{sec:methods_original_formulation}). The relative importance of each term is specified by weighting factors $\lambda_{rew}$, $\lambda_{SM}$, and $\lambda_{rxn}$, respectively. %
While minimizing reaction penalties encourages the selection of common intermediates, the objective is penalized only once for a reaction that may be used to synthesize multiple candidates. Further, even if the total number of high-risk reaction steps is the same between two candidate sets, it matters whether the high-risk reactions correspond to routes to the highest-value candidates. %

Analogous to Bayesian optimization acquisition functions that balance risk and reward through expectation values (e.g., expected improvement) \cite{garnett_bayesian_2023}, we introduce \er{},  
defined mathematically in Section \ref{sec:methods_expected_reward}. The expected reward associated with a single selected candidate is the product of its reward and the probabilities of success of every step in its synthetic route. The \er{} is obtained by summing over all selected compounds. In comparison to the weighted sum objective, \er{} penalizes uncertain reactions multiple times if used in multiple selected routes. While \er{} does not explicitly minimize the number of reaction steps or the cost of starting materials, budgets on starting material costs and the number of selected reactions can be incorporated as inequality constraints to encourage the selection of routes with shared building blocks and intermediates, respectively. This revised formulation results in a nonlinear program (Section \ref{sec:methods_expected_reward}) and requires nonlinear solvers for its maximization. %

\begin{figure}[ht]
    \centering
    \includegraphics{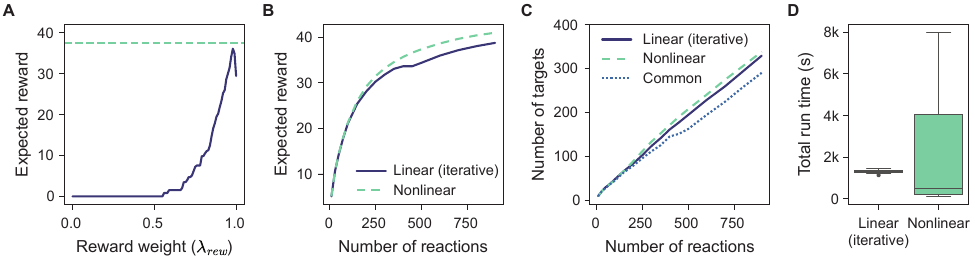}
    \caption{Development of a linear optimization strategy to maximize expected cumulative reward. (A) Taking the starting material weighting factor to be zero ($\lambda_{SM}=0$) and constraining the two remaining weights to sum to one ($\lambda_{rew}+\lambda_{rxn} = 1$), the cumulative expected reward achieved through direct nonlinear optimization (dashed) is nearly achieved for certain values of the reward weighting factor ($\lambda_{rew}$). Here, the maximum number of reactions is set to 500. (B) A linear optimization strategy that iteratively tunes $\lambda_{rew}$ leads to results nearly identical to the nonlinear optimization of expected reward under varied reaction constraints. (C) Both optimization approaches select a similar number of candidates, with a majority selected in both solutions (``Common''). (D) The total run time required by the linear strategy is longer on average but less variable.   %
    }
    \label{fig:linear}
\end{figure}

When constraining $\lambda_{SM}=0$ and $\lambda_{rxn} = 1-\lambda_{rew}$, we observe that certain values of $\lambda_{rew}$ in our weighted-sum objective yield solutions that are almost equivalent to that obtained when directly optimizing expected cumulative reward (Figure \ref{fig:linear}A). To efficiently identify these values of $\lambda_{rew}$, we implemented an iterative optimization scheme that maximizes \er{} with respect to $\lambda_{rew}$ %
(Section \ref{sec:methods_tuning}). 

We applied both optimization strategies (Sections \ref{sec:methods_expected_reward} and \ref{sec:methods_tuning}) to downselect compounds and syntheses from the candidate antibiotics proposed by AutoMolDesigner \cite{shen_automoldesigner_2024}. Both arrive at solutions with similar values of \er{} and select many candidates in common (Figure \ref{fig:linear}B and C). The time required to perform the iterative linear optimization is longer on average, although the nonlinear solve times exhibit higher variability (Figure \ref{fig:linear}D). Because SPARROW's implementation of the nonlinear optimization relies on the commercial solver Gurobi \cite{noauthor_gurobi_2024} and nonlinear solvers in general are not guaranteed to converge to a global optimum \cite{arora_methods_1994}, we use the linear formulation for all following analysis. 

\subsection{Encouraging molecular diversity with cluster representation}
\label{sec:diversity}
Diversity is often a primary criterion in the downselection of compounds in molecular design cycles \cite{gorse_molecular_1999, gorse_functional_2000, logan_discovery_2024}. For example, it is common in virtual screening to cluster top-performers and select cluster representatives \cite{kaplan_bespoke_2022, liu_structure-based_2024} to reduce the risk of an entire chemical series failing to experimentally validate and/or reveal the same liabilities in secondary property assays. %
Promoting diversity is also relevant when selecting from libraries proposed by goal-directed generative models, which can have a propensity to generate highly similar compounds \cite{renz_diverse_2024}. The definition of ``diversity'' varies across applications and design projects; it could be defined in terms of structural similarity using conventional structural fingerprint definitions, the presence of specific scaffolds \cite{uguen_buildcoupletransform_2022}, predicted interactions with a protein target \cite{makara_measuring_2001, deng_structural_2004, bouysset_prolif_2021}, or other hypotheses relating structure to function \cite{bradley_rapid_2000, tafi_pharmacophore_2006}. 

We define a cluster-based formulation that enables SPARROW to flexibly optimize various types of diversity (Section \ref{sec:methods_cluster_formulation}). We assume that each candidate is assigned to one or more clusters. %
We add a term to the scalarized objective function equal to the number of clusters represented by the selected candidate set (Section \ref{sec:methods_cluster_formulation}), encouraging the selection of a diverse batch that comprises many clusters. The weighting factor associated with this term, $\lambda_{div}$, assigns the relative importance of diversity. Diversity may alternatively be imposed as a constraint on the number of clusters represented by the selected compounds, as proposed by \citet{briem_diversity-oriented_2024}.

\begin{figure}[t]
    \centering
    \includegraphics{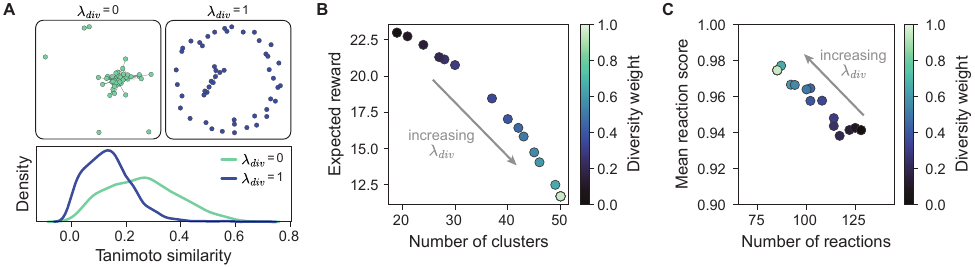}
    \caption{Cluster-based formulation improves diversity of selected candidates. (A) Compound selections are visualized as a network wherein nodes represent acquired compounds and edges are drawn between pairs with Tanimoto similarity $>$ 0.35. Nodes are positioned using the Fruchterman-Reingold force-directed algorithm. \citep{fruchterman_graph_1991} The greater spread of nodes indicates higher diversity, which can also be visualized as the shift in the distribution of Tanimoto similarity scores for all pairs of selections. (B) Increasing the diversity weight ($\lambda_{div}$) increases the number of clusters represented in the solution, although this results in a decrease in expected reward. (C) Increasing $\lambda_{div}$ generally decreases the number of selected reactions and increases the mean reaction score, demonstrating that diversity can be obtained without substantially more reaction steps and without reactions that have a high risk of failure. }
    \label{fig:diversity}
\end{figure}

We demonstrate this formulation on the same set of candidate antibiotics generated by AutoMolDesigner \cite{shen_automoldesigner_2024}, defining clusters using the Butina clustering algorithm on count Morgan fingerprints \cite{landrum_rdkit_2024} (Section \ref{sec:methods_cluster_formulation}). We vary the diversity weighting factor $\lambda_{div}$ while constraining the number of candidate selections to 50. We visualize the change in the diversity of SPARROW's selections as both networks with edges drawn between compound pairs with Tanimoto similarity > 0.35 and histograms of pairwise Tanimoto similarity of count Morgan fingerprints for all pairs of selected compounds (Figure \ref{fig:diversity}A). %
Figure \ref{fig:diversity}B depicts how the Pareto front of diversity and expected reward can be traversed by varying $\lambda_{div}$. %
Despite the inherent trade-off between maximizing diversity and selecting compounds with the highest reward values, compound diversity can still be achieved without substantially greater numbers of reaction steps and without riskier reaction steps (Figure \ref{fig:diversity}C). In this particular case, prioritizing diversity over individual reward increases reaction scores and reduces the number of selected reactions, indicating that the relative benefit of synthesizing very high-reward compounds with riskier reactions is diminished by the preference for diversity. 

\subsection{Promoting the selection of batches that can be prepared using parallel chemistry}
\label{sec:parallel_chemistry}

\begin{figure}
    \centering
    \includegraphics{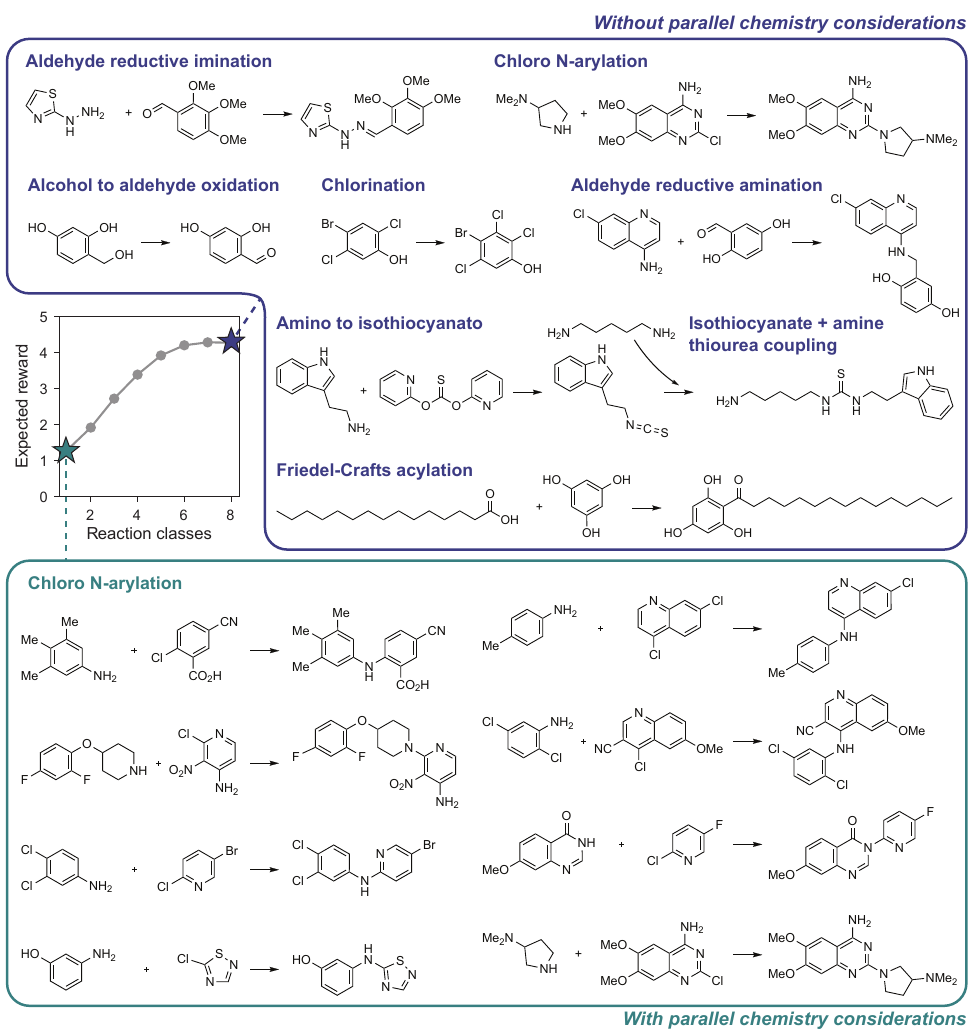}
    \caption{Constraining the number of selected reaction classes improves amenability to parallel chemistry. Limiting the selection to a single NameRxn \cite{noauthor_hazelnut_nodate} class leads to only chloro N-arylation reactions. This set of syntheses is amenable to parallel chemistry, although the constraint leads to a lower overall expected reward. In contrast, allowing up to eight NameRxn \cite{noauthor_hazelnut_nodate} classes leads to syntheses that are perhaps less amenable to parallel chemistry but with a higher expected reward. This decrease in expected reward can be attributed to a decrease in both the likelihood of success of the selected reactions and the reward values of selected candidates (Table \ref{tab:parallel_table}). All reaction class labels were assigned with NameRxn \cite{noauthor_hazelnut_nodate}.}
    \label{fig:parallel-chemistry}
\end{figure}

Parallel synthesis can substantially reduce the the effort required to synthesize a batch of compounds \cite{labaudiniere_rprs_1998, schneider_automating_2018} and support the rapid discovery of promising compounds \cite{cavallaro_discovery_2012, cheung_parallel_2016, sadybekov_computational_2023}. 
Here, we extend SPARROW to include a heuristic %
that encourages the selection of parallelizable reactions. 
We assume that reactions may be classified (e.g., with SMARTS patterns \cite{noauthor_smarts_2022}) and that reactions in a single class can be run in parallel.
While this assumption may fail when specific substrates require different conditions or when the reaction sequence prevents parallelism, constraining the number of reaction classes can still favor routes that are in general more amenable to parallel synthesis. 
We implement this as an inequality constraint in SPARROW's formulation to limit the number of distinct reaction classes selected (Section \ref{sec:methods_cluster_formulation}). 

We apply this algorithm to the same case study, constraining the selection to a maximum of eight reactions (to keep the number of selections small for visualization purposes). We define reaction classes with NameRxn \cite{noauthor_hazelnut_nodate}, although a more restrictive set of SMARTS strings could be used to define reaction types compatible with a particular parallel chemistry platform. SPARROW's selections with and without parallel chemistry considerations are shown in Figure \ref{fig:parallel-chemistry}. %
SPARROW identified the chloro N-arylation transformation as the most effective for parallel synthesis of high-reward compounds in this candidate set. %
This example highlights the commonly encountered trade-off between expected reward and amenability to parallel chemistry; limiting the number of reaction classes impacts both the reward values of selected compounds and the likelihood of success of selected reactions (Table \ref{tab:parallel_table}).

\subsection{Simultaneously encouraging batch diversity and amenability of designs to  parallel chemistry}

\begin{figure}
    \centering
    \includegraphics{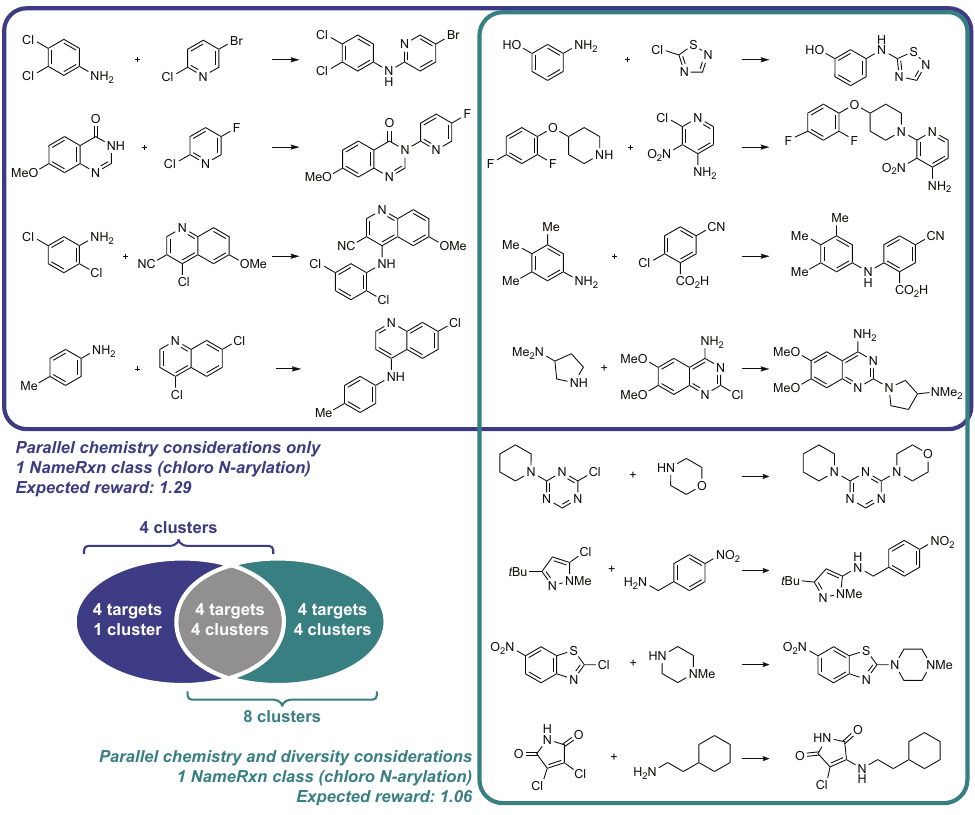}
    \caption{Combining SPARROW capabilities to synthesize diverse batches with parallel chemistry. When SPARROW's formulation considers parallel chemistry without diversity in this example, only four clusters are represented by the eight candidate compounds selected. When both diversity and amenability to parallel chemistry are incorporated into SPARROW's optimization, the selections span eight distinct clusters. In both cases, all reactions are classified as chloro N-arylation reactions by NameRxn \cite{noauthor_hazelnut_nodate}. Selected compound SMILES, rewards and cluster assignments, as well as reaction likelihood scores, are provided in Table \ref{tab:combining_table}. }
    \label{fig:combining}
\end{figure}

Libraries synthesized with parallel chemistry require careful expert design to ensure sufficient coverage of chemical space \cite{nielsen_towards_2008, follmann_approach_2019}. For example, DNA-encoded libraries (DELs) provide a high-throughput approach to testing many compounds synthesized in parallel, but their diversity can be limited depending on the details of chemistry and %
building block selection \cite{satz_dna_2015, lenci_diversity-oriented_2021}. 
In general, there is a natural trade-off between structural diversity and amenability to parallel chemistry, although this can be mitigated with access to sufficiently diverse building block collections \cite{goldberg_designing_2015, tomberg_can_2020}. Here, we demonstrate how SPARROW can manage this trade-off in the design of diverse libraries amenable to parallel synthesis. %

Continuing with the same candidate set and clusters defined previously, the products of the chloro N-arylation reactions shown in Figure \ref{fig:parallel-chemistry} represent four distinct structural clusters. We combined the diversity and parallel chemistry formulations (Section \ref{sec:methods_combining}) and applied this strategy to SPARROW's downselection with a maximum of eight reactions. %
As demonstrated by the results in Figure \ref{fig:combining}, SPARROW identifies a new set of reactions and targets that represents eight distinct clusters while maintaining amenability to parallel chemistry. 

\section{Conclusion}

The downselection of compounds for synthesis and testing is a key decision that drives molecular design cycles. Algorithms that support this decision-making procedure promise to improve the efficiency of resource utilization in molecular discovery, support the transition to autonomous workflows, and scale expert chemist intuition to large chemical libraries. The development of such algorithms is challenged by the complexity of compound downselection; synthetic route planning, reaction feasibility, compound utility, amenability to parallel chemistry, and compound diversity are just a few potential considerations relevant to compound selection. 

In this work, we have extended the capabilities of the downselection framework SPARROW \cite{fromer_algorithmic_2024} to capture additional complexities often encountered in molecular cycles. Specifically, we enable SPARROW to (1) consider compound utilities and synthesis risk in a more principled manner by optimizing \er{}, (2) improve compound diversity with flexibility in the definition of clusters including and beyond structural similarity, and (3) select synthetic routes that are amenable to parallel chemistry. As demonstrated on a set of candidate antibiotics proposed by AutoMolDesigner \cite{shen_automoldesigner_2024}, these mathematical formulations enable the use of SPARROW when diversity must be prioritized and when high-throughput synthesis is preferred. Further, they more precisely capture the expert intuition commonly used to design and select compounds. 

SPARROW is publicly available at \url{https://github.com/coleygroup/sparrow} to support its use by the molecular design community. While we demonstrated SPARROW on a set of approximately 10k de novo designs, SPARROW may also be applied to smaller or larger sets of compounds designed by experts, extracted from make-on-demand libraries, or proposed by generative models. Additionally, clusters may be defined by project-specific criteria (e.g., interactions with protein targets or presence of specific fragments), and reactions may be classified according to custom SMARTS patterns instead of NameRxn. These design choices are intended to flexibly support SPARROW's application to diverse molecular design campaigns. %

\section{Methods}
\label{section:methods}

\subsection{Synthetic pathway generation}
\label{sec:methods_network_construction}

The Monte Carlo tree search retrosynthesis model in ASKCOS \cite{tu_askcos_2025} was used to generate retrosynthetic pathways to each of the 9,621 candidate \emph{Staphylococcus aureus} antibiotics generated by AutoMolDesigner \cite{shen_automoldesigner_2024}. The retrosynthetic search was run on nodes with four CPUs with an expansion time of 60 seconds, a branching factor of 25, and a maximum depth of 6. All other search parameters were set to ASKCOS defaults. ASKCOS identified synthetic routes to 3,517 of the candidates. AutoMolDesigner is not a synthesizability-constrained generative model, so the success rate of 37\% is reasonable. In total, the full set of hypothetical pathways to all candidate molecules included 380,650 unique reactions.

Reaction conditions were proposed by the Reaxys-trained neural network condition recommendation model \cite{gao_using_2018} in ASKCOS. The probability of reaction success was estimated using a forward predictor model \cite{coley_graph-convolutional_2019} %
in ASKCOS trained on the Pistachio dataset \cite{noauthor_pistachio--nextmove_nodate}. 

We then pruned the reaction network to reduce the total number of decision variables in the optimization problems (Section \ref{sec:methods_formulations}), alleviating memory issues and reducing solve times. Any reaction with a predicted likelihood of success less than 0.01 was removed from the network. Next, any compound or reaction node that was not within a distance of 16 from any candidate was removed, as calculated with the \texttt{shortest\_path\_length} function on a \texttt{networkx} graph instance of the reaction network \cite{hagberg_exploring_2008}. These pruning steps reduced the size of the network to 261,140 reaction nodes and 121,514 compound nodes. 3,254 candidates remained in the network. Compounds were queried against Enamine's building block database \cite{noauthor_enamine_nodate} to determine buyability. 

\subsection{Optimization procedures}
All linear formulations were implemented with PuLP \cite{dunning_pulp_2011} and solved with the coin-or branch and cut solver \cite{forrest_coin-orcbc_2023}. Direct optimization of \er{} were performed with Gurobi \cite{noauthor_gurobi_2024}. SPARROW was run on nodes with 48 CPUs on the MIT Supercloud \cite{reuther_interactive_2018}. 

All optimization problem formulations are defined mathematically in Section \ref{sec:methods_formulations}.

\section*{Data and Software Availability} %
The repository at \url{https://github.com/coleygroup/sparrow} includes all code to run SPARROW and directions for its installation and use. Directions and data for reproducing the results in this work specifically are provided at \url{https://github.com/coleygroup/sparrow/tree/main/examples/automoldesigner}. This folder also contains compound buyability and reaction likelihood of success scores. A NameRxn executable is required to reproduce SPARROW runs constraining reaction class representation, and a Gurobi license is required to reproduce SPARROW runs that directly optimize \er{}.

\section*{Conflicts of Interest}
There are no conflicts to declare. 

\section*{Acknowledgment}

This work was supported by the Machine Learning for Pharmaceutical Discovery and Synthesis consortium. 
J.C.F. received additional support from the National Science Foundation Graduate Research Fellowship under Grant No. 2141064. A.D.V. acknowledges support from the MIT Undergraduate Research Opportunities Program. 
The authors acknowledge the MIT SuperCloud %
and the Lincoln Laboratory Supercomputing Center for providing resources that have contributed to the research results reported within this paper. We are grateful to Marco Stenta, Marta Pasquini, and Tobias Ziegler for participating in valuable discussions that guided the development of SPARROW. 

\clearpage 
\bibliography{references.bib} 

\renewcommand*{\thesection}{S\arabic{section}}
\renewcommand*{\thetable}{S\arabic{table}}
\renewcommand*{\thefigure}{S\arabic{figure}}
\setcounter{section}{0}
\setcounter{table}{0}
\setcounter{figure}{0}

\clearpage 
\begin{suppinfo}
\section{Mathematical formulations}
\label{sec:methods_formulations}

\begin{table}[H]
\small 
    \centering
    \renewcommand{\arraystretch}{1.2}
    \begin{tabular}{|c|c|l|} \hline
        & Symbol   & Description \\ \hline 
        \multirow{6}{*}{\rotatebox[origin=c]{90}{Variables}} &$\cpdpicked{\cpdind}$ & Binary variable indicating whether compound $\cpdind$ is in any selected route \\
        &$\rxnpicked{\rxnind}$ & Binary variable indicating whether reaction $\rxnind$ is in any selected route \\ 
        &$\cpdfor{\cpdind}{\tarind}$ & Binary variable indicating whether compound $\cpdind$ is in the selected route to target $\tarind$ \\
        &$\rxnfor{\rxnind}{\tarind}$ & Binary variable indicating whether reaction $\rxnind$ is in the selected route to target $\tarind$ \\
        &$\clusterpicked_\clusind$ & Binary variable indicating whether cluster $\clusind$ is represented by the selected targets \\ 
        & $\classpicked_\classind$ & Binary variable indicating whether any selected reaction falls into class $\classind$ \\
        \hline 
        \multirow{4}{*}{\rotatebox[origin=c]{90}{Values}} &$\reward_\tarind$ & Reward value for target $\tarind$ \\
        &$\rpenalty_\rxnind$ & Penalty associated with selected reaction $\rxnind$ \\
        &$\smcost_\rxnind$ & Cost of purchasing the compound generated by dummy reaction $\rxnind$ \\
        &$\likelihoodscore_\rxnind$ & Likelihood of reaction success for reaction $\rxnind$ \\
        \hline
        \multirow{12}{*}{\rotatebox[origin=c]{90}{Sets}} &$\sms$ & The set of dummy reaction indices that generate purchasable compounds \\ 
        &$\allrxns$ & The set of all candidate reaction indices \\ 
        &$\allcpds$ & The set of all compound indices in the network \\ 
        &$\targets$ & The subset of compound indices corresponding to candidates (i.e. targets) \\
        &$\cycle$ & The set of reaction and compound indices that form a cycle \\ 
        &$\compoundsincluster{\clusind}$ & The set of compound indices that are in cluster $\clusind$ \\ 
        &$\clusters$ & The set of compound clusters \\
        &$\rxnclasses$ & The set of reaction classes \\
        &$\allrxns_\classind$ & The subset of reaction indices that fall into reaction class $\classind$\\
        &$\allrxns_\tarind$ & The subset of reaction indices within a distance of $\prunedistance$ to target $\tarind$\\
        &$\allcpds_\tarind$ & The subset of compound indices within a distance of $\prunedistance$ to target $\tarind$\\
        &$\targets_\rxnind$ & The subset of candidate indices within a distance of $\prunedistance$ from reaction $\rxnind$ \\
        &$\targets_\cpdind$ &  The subset of candidate indices within a distance of $\prunedistance$ from compound $\cpdind$ \\
        \hline
        \multirow{4}{*}{\rotatebox[origin=c]{90}{Weights}} & $\lambda_{rew}$ & Weighting factor for maximizing compound reward \\ 
        &$\lambda_{rxn}$ & Weighting factor for minimizing reaction penalties \\ 
        &$\lambda_{SM}$ & Weighting factor for minimizing starting material costs \\ 
        &$\lambda_{div}$ & Weighting factor for maximizing diversity \\ \hline
        \multirow{5}{*}{\rotatebox[origin=c]{90}{Constraints}} & $\maxrxns$ & Maximum number of reactions \\
        & $\maxtars$ & Maximum number of candidates selected \\
        & $\budget$ & Starting material budget \\
        & $\maxclasses$ & Maximum number of reaction classes \\ 
        & $\prunedistance$ & Pruning distance for nonlinear optimization \\\hline 
    \end{tabular}
    \caption{Notation for all mathematical formulations defined below. }
    \label{tab:notation}
\end{table}

\clearpage
\subsection{Scalarized objective in initial SPARROW formulation \cite{fromer_algorithmic_2024}}
\label{sec:methods_original_formulation}
The optimization problem solved by the initial formulation of SPARROW \cite{fromer_algorithmic_2024} can be defined as the following: 

{\setstretch{1}
\begin{align*}
   & & \argmax_{\boldsymbol{\rr, \, \cc}} \;
     \lambda_{rew} \sum_{\tarind \in \targets}  & \reward_\tarind  \cpdpicked{\tarind}-
    \lambda_{rxn} \sum_{\rxnind \in \allrxns} \rpenalty_{\rxnind} \rxnpicked{\rxnind} -
    \lambda_{SM} \sum_{\rxnind \in \sms }  \smcost_\rxnind \rxnpicked{\rxnind} \\ 
    & s.t. & & \\
    &  & \cpdpicked{\cpdind} \geq \rxnpicked{\rxnind}  \: \: & \text{for all compounds $\cpdind$ that are reactants of reaction $\rxnind$,} \\
    & & & \text{for all reactions $\rxnind$} \\ 
    & & \sum_{\rxnind \in \parents_\cpdind} \rxnpicked{\rxnind} \geq \cpdpicked{\cpdind} \: \: & \text{for all compounds $\cpdind$} \\
    & & \sum_{\rxnind \in \cycle} \rxnpicked{\rxnind} < |\cycle| \: \: & \text{for all cycles $\cycle$}  
\end{align*}}

The scalarized objective function is comprised of three terms that (1) maximize the reward of selected candidates, (2) minimize penalties associated with low-confidence reactionss, and (3) minimize starting material costs. Weighting factors $\lambda$ determine the relative importance of each objective. Reaction penalties are defined as
\begin{align*}
    \rpenalty_\rxnind = \max (\frac{1}{\likelihoodscore_\rxnind}, 20).
\end{align*}
Constraints ensure that all reactants of a selected reaction are selected and that at least one parent reaction of any selected compound is selected. Finally, all reactions that form a cycle in the retrosynthetic graph cannot be selected simultaneously. 

The following optional constraints may also be included: 
\begin{align*}
     \sum_{\rxnind \in \allrxns} \rxnpicked{\rxnind} & \leq  \: \maxrxns \: &  & \text{Maximum number of reactions (optional)}  \\
     \sum_{\rxnind \in \sms }  \smcost_\rxnind \rxnpicked{\rxnind} & \leq  \: \budget & & \text{Budget on starting materials (optional)} \\
     \sum_{\tarind \in \targets} \cpdpicked{\tarind} & \leq  \: \maxtars \: & & \text{Maximum number of targets (optional)} 
\end{align*}

\clearpage
\subsection{Direct optimization of expected reward}
\label{sec:methods_expected_reward}
Expected reward associated with a single compound is that compounds reward multiplied by the likelihood of success for the target's \emph{entire} synthetic route. Synthetic route success likelihood is equal to $\prod_{\rxnind \in \allrxns_{\tarind}} \likelihoodscore_\rxnind$, or the product of the likelihood scores for all reactions in the synthetic route to target $\tarind$. Using the binary decision variable $\rxnfor{\rxnind}{\tarind}$, which defines whether reaction $\rxnind$ is in the selected route to target $\tarind$, expected reward for selected target $\tarind$ is equal to $\prod_{\rxnind \in \allrxns} (\likelihoodscore_\rxnind)^{\rxnfor{\rxnind}{\tarind}}$ where $\allrxns$ represents the set of \emph{all} reactions in the network. 

This formulation requires additional decision variables defining whether a reaction or compound is used in the synthetic route to a particular target, leading to additional constraints. Nonetheless, these constraints are conceptually equivalent to those in the linear formulation (Section \ref{sec:methods_original_formulation}), extended to the additional decision variables. Constraints also ensure that any reaction or compound that is selected in the route to any target is selected in general. 

This leads to the following formulation for maximizing expected cumulative reward, the sum of expected reward for all selected targets: 
\clearpage
\begin{equation*}
    \argmax_{\boldsymbol{\rr, \, \cc, \, \rfor, \, \cfor}} \;
     \;
    \sum_{ \tarind \in \targets} \left( \reward_{\tarind} 
 \cpdpicked{\tarind} 
    \prod_{\rxnind \in \allrxns} (\likelihoodscore_{\rxnind})^{\rxnfor{\rxnind}{\tarind}}
    \right) 
\end{equation*}
$s.t.$
{\setstretch{1}
\begin{align*}
    \cpdfor{\tarind}{\tarind} = & \: \cpdpicked{\tarind} \: \  & \text{for} & \text{ all targets $\tarind$}    \\
    \cpdpicked{\cpdind} \geq & \: \rxnpicked{\rxnind}  \: \: & \text{for} & \text{ all compounds $\cpdind$ that are reactants of reaction $\rxnind$,} \\
     && \text{for} & \text{ all reactions $\rxnind$} \\ 
    \cpdfor{\cpdind}{\tarind} \geq & \: \rxnfor{\rxnind}{\tarind}  \: \: & \text{for} & \text{ all compounds $\cpdind$ that are reactants of reaction $\rxnind$,} \\
     && \text{for} & \text{ all reactions $\rxnind$} \\ 
     \sum_{\rxnind \in \parents_\cpdind} \rxnfor{\rxnind}{\tarind} \geq & \: \cpdfor{\cpdind}{\tarind} \: \: & \text{for} & \text{ all targets $\tarind$ and all compounds $\rxnind$} \\ 
     \sum_{\rxnind \in \parents_\cpdind} \rxnpicked{\rxnind} \geq & \: \cpdpicked{\cpdind} \: \: & \text{for} & \text{ all compounds $\cpdind$} \\
     \sum_{\rxnind \in \cycle} \rxnpicked{\rxnind} < & \: |\cycle| \: \:  & \text{for} & \text{ all cycles $\cycle$} \\
     \frac{1}{|\targets|} \sum_{\tarind \in \targets} \rxnfor{\rxnind}{\tarind} \leq  & \: \rxnpicked{\rxnind} \: \ & \text{for} & \text{ all reactions $\rxnind$} \\ 
    \sum_{\tarind \in \targets} \rxnfor{\rxnind}{\tarind} \geq  & \: \rxnpicked{\rxnind} \: \ & \text{for} & \text{ all reactions $\rxnind$} \\ 
    \frac{1}{|\targets|} \sum_{\tarind \in \targets} \cpdfor{\cpdind}{\tarind} \leq & \: \cpdpicked{\cpdind} \: \  & \text{for} & \text{ all compounds $\cpdind$}    \\
    \sum_{\tarind \in \targets} \cpdfor{\cpdind}{\tarind} \geq  & \: \cpdpicked{\cpdind} \: \ & \text{for} & \text{ all compounds $\cpdind$} \\ 
    \frac{1}{|\allrxns|} \sum_{\rxnind \in \allrxns} \rxnfor{\rxnind}{\tarind} \leq & \: \cpdpicked{\tarind} \: \  & \text{for} & \text{ all targets $\tarind$}    \\
    \frac{1}{|\allcpds|} \sum_{\cpdind \in \allcpds} \cpdfor{\cpdind}{\tarind} \leq & \: \cpdpicked{\tarind} \: \  & \text{for} & \text{ all targets $\tarind$}    \\
     \sum_{\rxnind \in \allrxns} \rxnpicked{\rxnind} \leq & \: \maxrxns \: &  \text{Ma} &\text{ximum number of reactions (optional)}  \\
     \sum_{\rxnind \in \sms }  \smcost_\rxnind \rxnpicked{\rxnind} \leq & \: \budget & \text{Bu} & \text{dget on starting materials (optional)} \\
     \sum_{\tarind \in \targets} \cpdpicked{\tarind} \leq & \: \maxtars \: &  \text{Ma} &\text{ximum number of targets (optional)} 
\end{align*}}

For large retrosynthetic networks ($O(10^5)$ reactions) and large candidate sets ($O(10^3)$ candidates), this may correspond to to optimization problems with $O(10^8)$ decision variables. Because we observed memory issues when solving this optimization for the AutoMolDesigner \cite{shen_automoldesigner_2024} case study, we also implemented a slightly modified version of this optimization problem that assumes selected synthetic routes to be below a certain threshold length. We define a pruning distance $\prunedistance$ which is double the threshold synthetic route length. In this formulation, the variables $\rxnfor{\rxnind}{\tarind}$ are only defined for reactions $\rxnind$ that are within a distance of $\prunedistance$ from candidate node $\tarind$. The  objective function and constraints for this pruned optimization are shown below. All other constraints remain identical to those defined above.  

\begin{equation*}
    \argmax_{\boldsymbol{\rr, \, \cc, \, \rfor, \, \cfor}} \;
     \;
    \sum_{ \tarind \in \targets} \left( \reward_{\tarind} 
 \cpdpicked{\tarind} 
    \prod_{\rxnind \in \allrxns_{\tarind}} (\likelihoodscore_{\rxnind})^{\rxnfor{\rxnind}{\tarind}}
    \right) 
\end{equation*}
$s.t.$
{\setstretch{1}
\begin{align*}
    \cpdfor{\cpdind}{\tarind} \geq & \: \rxnfor{\rxnind}{\tarind}  \: \: & \text{for} & \text{ all compounds $\cpdind$ that are reactants of reaction $\rxnind$,} \\
     && \text{for} & \text{ all reactions $\rxnind$ in $\allrxns_\tarind$ and compounds $\cpdind$ in $\allcpds_\tarind$} \\ 
     \sum_{\rxnind \in \parents_\cpdind} \rxnfor{\rxnind}{\tarind} \geq & \: \cpdfor{\cpdind}{\tarind} \: \: & \text{for} & \text{ all targets $\tarind$} \\
     && \text{for} & \text{ all compounds $\cpdind$ in $\allcpds_\tarind$ and reactions $\rxnind$ in $\allrxns_\tarind$}\\ 
     \frac{1}{|\targets_\rxnind|} \sum_{\tarind \in \targets_\rxnind} \rxnfor{\rxnind}{\tarind} \leq  & \: \rxnpicked{\rxnind} \: \ & \text{for} & \text{ all reactions $\rxnind$} \\ 
    \sum_{\tarind \in \targets_\rxnind} \rxnfor{\rxnind}{\tarind} \geq  & \: \rxnpicked{\rxnind} \: \ & \text{for} & \text{ all reactions $\rxnind$} \\ 
    \frac{1}{|\targets_\cpdind|} \sum_{\tarind \in \targets_\cpdind} \cpdfor{\cpdind}{\tarind} \leq & \: \cpdpicked{\cpdind} \: \  & \text{for} & \text{ all compounds $\cpdind$}    \\
    \sum_{\tarind \in \targets_\cpdind} \cpdfor{\cpdind}{\tarind} \geq  & \: \cpdpicked{\cpdind} \: \ & \text{for} & \text{ all compounds $\cpdind$} \\ 
    \frac{1}{|\allrxns_\tarind|} \sum_{\rxnind \in \allrxns_\tarind} \rxnfor{\rxnind}{\tarind} \leq & \: \cpdpicked{\tarind} \: \  & \text{for} & \text{ all targets $\tarind$}    \\
    \frac{1}{|\allcpds_\tarind|} \sum_{\cpdind \in \allcpds_\tarind} \cpdfor{\cpdind}{\tarind} \leq & \: \cpdpicked{\tarind} \: \  & \text{for} & \text{ all targets $\tarind$}   
\end{align*}}
The results shown in this work for directly optimizing expected reward (Figure \ref{fig:linear}) use this pruned formulation with $\prunedistance=16$. SPARROW requires a Gurobi \cite{noauthor_gurobi_2024} license to directly optimize expected reward with or without pruning. 

\clearpage
\subsection{Weighting factor tuning to optimize expected reward}
\label{sec:methods_tuning}
First, we set $\lambda_{SM}=0$ in the formulation described in Section \ref{sec:methods_original_formulation} and require $\lambda_{rew}+\lambda_{rxn}=1$. We allow optional constraints on starting material budget, total reaction count, and total number of targets. This results in the following objective function and constraints: 
\begin{equation*}
    \boldsymbol{r^*}(\lambda_{rew}), \boldsymbol{c^*}(\lambda_{rew}) = \argmax_{\boldsymbol{\rr, \, \cc}} \;
     \lambda_{rew} \sum_{\tarind \in \targets}  \reward_\tarind  \cpdpicked{\tarind}-
    (1-\lambda_{rew}) \sum_{\rxnind \in \allrxns} \rpenalty_{\rxnind} \rxnpicked{\rxnind}.
\end{equation*}
$s.t.$
{\setstretch{1}
\begin{align*}
    \cpdpicked{\cpdind} \geq & \: \rxnpicked{\rxnind}  \: \: & & \text{for all compounds $\cpdind$ that are reactants of reaction $\rxnind$,} \\
    & & & \text{for all reactions $\rxnind$} \\ 
    \sum_{\rxnind \in \parents_\cpdind} \rxnpicked{\rxnind} \geq & \: \cpdpicked{\cpdind} \: \: & & \text{for all compounds $\cpdind$} \\
    \sum_{\rxnind \in \cycle} \rxnpicked{\rxnind} < & \: |\cycle| \: \: & & \text{for all cycles $\cycle$}  \\
     \sum_{\rxnind \in \allrxns} \rxnpicked{\rxnind} \leq & \:  \maxrxns \: &  & \text{Maximum number of reactions (optional)}  \\
     \sum_{\rxnind \in \sms }  \smcost_\rxnind \rxnpicked{\rxnind} \leq & \: \budget & & \text{Budget on starting materials (optional)} \\
     \sum_{\tarind \in \targets} \cpdpicked{\tarind} \leq & \: \maxtars \: &&   \text{Maximum number of targets (optional)} 
\end{align*}}

The \er{} of the resulting solution, $\text{ER}(\boldsymbol{r^*}, \boldsymbol{c^*})$, can be maximized with respect to $\lambda_{rew}$ as the following optimization problem: 

{\setstretch{1}
\begin{equation*}
     \lambda_{rew}^* = \argmax_{\lambda_{rew}} \;
     \text{ER}(\boldsymbol{r^*}(\lambda_{rew}), \boldsymbol{c^*}(\lambda_{rew}))
\end{equation*}
\begin{align*}
    s.t. \: \: & & & \\
    & \lambda_{rew} \leq  \; 1-10^{-5} \\
     & \lambda_{rew} \geq \; 10^{-5}  
\end{align*}}

We implement this outer optimization loop using the \texttt{gp\_minimize} function in Scikit-learn \cite{pedregosa_scikit-learn_2011}. This function uses Bayesian optimization to iteratively optimize \er{}. We use an expected improvement acquisition function and initialize the optimization with two randomly selected values of $\lambda_{rew}$ (with a random seed of 0 for reproducibility). We allow a maximum of 20 iterations. Because this is a single parameter optimization, it would also be possible to evaluate a finite set of possibilities (e.g., $\lambda_{rew}$ = [0:0.05:1]). We postulate that the model-guided approach will identify more optimal solutions within a limited budget because it is not limited to a discrete set of possibilities for $\lambda_{rew}$. 

\clearpage
\subsection{Cluster formulation for diversity}
\label{sec:methods_cluster_formulation}

This formulation promotes the selection of diverse batches by maximizing the number of clusters represented by the select candidates. While SPARROW accommodates manually defined clusters, the clusters generated in this work were computed using the Butina clustering algorithm on count Morgan fingerprints as implemented in RDKit \cite{landrum_rdkit_2024}. A distance threshold of 0.8 was used, and the number of points was set equal to the number of fingerprints being clustered. 

Using the notation in Table \ref{tab:notation}, the optimization problem can be defined as: 

{\setstretch{1}
\begin{align*}
     \argmax_{\boldsymbol{\rr, \, \cc}} \;
     \lambda_{rew}^* \sum_{\tarind \in \targets} \reward_\tarind  \cpdpicked{\tarind}-
    (1-\lambda_{rew}^*) \sum_{\rxnind \in \allrxns} \rpenalty_{\rxnind} \rxnpicked{\rxnind} +
    \lambda_{div} \sum_{\clusind \in \clusters }  \clusterpicked_{\clusind}
\end{align*}
$s.t.$
\begin{align*}
 \cpdpicked{\cpdind} & \geq \rxnpicked{\rxnind}  \: \: & & \text{for all compounds $\cpdind$ that are reactants of reaction $\rxnind$,} \\
 & & & \text{for all reactions $\rxnind$} \\ 
 \sum_{\rxnind \in \parents_\cpdind} \rxnpicked{\rxnind} & \geq \cpdpicked{\cpdind} \: \: & & \text{for all compounds $\cpdind$} \\
 \sum_{\rxnind \in \cycle} \rxnpicked{\rxnind} & < |\cycle| \: \:  & & \text{for all cycles $\cycle$}  \\
 \sum_{\cpdind \in \compoundsincluster{\clusind}} \cpdpicked{\cpdind} & \geq \clusterpicked_\clusind \: \: & & \text{for all clusters $\clusind$}\\
     \sum_{\rxnind \in \allrxns} \rxnpicked{\rxnind} & \leq  \: \maxrxns \: &  & \text{Maximum number of reactions (optional)}  \\
     \sum_{\rxnind \in \sms }  \smcost_\rxnind \rxnpicked{\rxnind} & \leq  \: \budget & & \text{Budget on starting materials (optional)} \\
     \sum_{\tarind \in \targets} \cpdpicked{\tarind} & \leq  \: \maxtars \: & & \text{Maximum number of targets (optional)} 
\end{align*}}

First, we set $\lambda_{div}=0$ and identify $\lambda_{rew}^*$ according to the procedure in Section \ref{sec:methods_tuning}. We then incorporate positive values of $\lambda_{div}$ to improve the diversity of selections while holding $\lambda_{rew}^*$ constant, leading to the results in Figure \ref{fig:diversity}. 

\clearpage
\subsection{Reaction classification for amenability to parallel chemistry}
\label{sec:methods_rxn_classes}
This formulation selects reactions amenable to parallel chemistry by constraining the number of distinct reaction classes represented by the selected reactions. While we use NameRxn \cite{noauthor_hazelnut_nodate} was used to classify reactions, SPARROW also accomodates manually defined reaction classes provided as a CSV file. We follow the scheme described in \ref{sec:methods_tuning} and define the additional binary variables $\classpicked_\classind$---whether reaction class $\classind$ is selected---and a corresponding constraint: 

{\setstretch{1}
\begin{equation*}
    \boldsymbol{r^*}(\lambda_{rew}), \boldsymbol{c^*}(\lambda_{rew}) = \argmax_{\boldsymbol{\rr, \, \cc}} \;
     \lambda_{rew} \sum_{\tarind \in \targets}  \reward_\tarind  \cpdpicked{\tarind}-
    (1-\lambda_{rew}) \sum_{\rxnind \in \allrxns} \rpenalty_{\rxnind} \rxnpicked{\rxnind}.
\end{equation*}
$s.t.$
\begin{align*}
    \cpdpicked{\cpdind} \geq & \: \rxnpicked{\rxnind}  \: \: & & \text{for all compounds $\cpdind$ that are reactants of reaction $\rxnind$,} \\
    & & & \text{for all reactions $\rxnind$} \\ 
    \sum_{\rxnind \in \parents_\cpdind} \rxnpicked{\rxnind} \geq & \: \cpdpicked{\cpdind} \: \: & & \text{for all compounds $\cpdind$} \\
    \sum_{\rxnind \in \cycle} \rxnpicked{\rxnind} < & \: |\cycle| \: \: & & \text{for all cycles $\cycle$}  \\
    \frac{1}{|\allrxns_\classind|} \sum_{\rxnind \in \allrxns_\classind} \rxnpicked{\rxnind} \leq & \: \classpicked_\classind  \: & & \text{for all classes $\classind$} \\
    \sum_{\classind \in \rxnclasses} \classpicked_\classind \leq & \: \maxclasses \: &&   \text{Maximum number of reaction classes} \\
     \sum_{\rxnind \in \allrxns} \rxnpicked{\rxnind} \leq & \:  \maxrxns \: &  & \text{Maximum number of reactions (optional)}  \\
     \sum_{\rxnind \in \sms }  \smcost_\rxnind \rxnpicked{\rxnind} \leq & \: \budget & & \text{Budget on starting materials (optional)} \\
     \sum_{\tarind \in \targets} \cpdpicked{\tarind} \leq & \: \maxtars \: &&   \text{Maximum number of targets (optional)} 
\end{align*}}

We continue by optimizing $\text{ER}(\boldsymbol{r^*}, \boldsymbol{c^*})$ with respect to $\lambda_{rew}$, while maintaining the reaction class constraint, using the same methods described in Section \ref{sec:methods_tuning}: 

{
\setstretch{1}
\begin{align*}
& \lambda_{rew}^* = \argmax_{\lambda_{rew}} \;
     \text{ER}(\boldsymbol{r^*}(\lambda_{rew}), \boldsymbol{c^*}(\lambda_{rew})) \\ 
    s.t. \: \: & & & \\
    & \lambda_{rew} \leq  \; 1-10^{-5} \\
     & \lambda_{rew} \geq \; 10^{-5}  
\end{align*}}

\clearpage
\subsection{Combining parallel chemistry and diversity considerations}
\label{sec:methods_combining}

First, we follow Section \ref{sec:methods_rxn_classes} to identify the $\lambda_{rew}^*$ that maximizes \er{} under the specified reaction class constraint. We then solve the following problem for positive values of $\lambda_{div}$ while holding $\lambda_{rew}^*$ constant:

{\setstretch{1}
\begin{align*}
     \argmax_{\boldsymbol{\rr, \, \cc}} \;
     \lambda_{rew}^* \sum_{\tarind \in \targets} \reward_\tarind  \cpdpicked{\tarind}-
    (1-\lambda_{rew}^*) \sum_{\rxnind \in \allrxns} \rpenalty_{\rxnind} \rxnpicked{\rxnind} +
    \lambda_{div} \sum_{\clusind \in \clusters }  \clusterpicked_{\clusind}
\end{align*}
$s.t.$
\begin{align*}
 \cpdpicked{\cpdind} & \geq \rxnpicked{\rxnind}  \: \: & & \text{for all compounds $\cpdind$ that are reactants of reaction $\rxnind$,} \\
 & & & \text{for all reactions $\rxnind$} \\ 
 \sum_{\rxnind \in \parents_\cpdind} \rxnpicked{\rxnind} & \geq \cpdpicked{\cpdind} \: \: & & \text{for all compounds $\cpdind$} \\
 \sum_{\rxnind \in \cycle} \rxnpicked{\rxnind} & < |\cycle| \: \:  & & \text{for all cycles $\cycle$}  \\
 \sum_{\cpdind \in \compoundsincluster{\clusind}} \cpdpicked{\cpdind} & \geq \clusterpicked_\clusind \: \: & & \text{for all clusters $\clusind$}\\
     \frac{1}{|\allrxns_\classind|} \sum_{\rxnind \in \allrxns_\classind} \rxnpicked{\rxnind} & \leq \: \classpicked_\classind  \: & & \text{for all classes $\classind$} \\
    \sum_{\classind \in \rxnclasses} \classpicked_\classind &\leq  \: \maxclasses \: & &   \text{Maximum number of reaction classes} \\
     \sum_{\rxnind \in \allrxns} \rxnpicked{\rxnind} & \leq  \: \maxrxns \: &  & \text{Maximum number of reactions (optional)}  \\
     \sum_{\rxnind \in \sms }  \smcost_\rxnind \rxnpicked{\rxnind} & \leq  \: \budget & & \text{Budget on starting materials (optional)} \\
     \sum_{\tarind \in \targets} \cpdpicked{\tarind} & \leq  \: \maxtars \: & & \text{Maximum number of targets (optional)} 
\end{align*}}

The results in Figure \ref{fig:combining} use $\lambda_{div}=0.1$, $\maxclasses=1$, and $\maxrxns=8$. Clusters were defined as detailed in Section \ref{sec:methods_cluster_formulation}. 

\clearpage
\section{Tables}

\bgroup
\def\arraystretch{0.8}
\Rotatebox{90}{
    \scriptsize
    \begin{tabular}{|l|l|l|l|l|}
    \hline
    \multrow{3}{Reactant SMILES} & \multrow{3}{Product SMILES} & \multrow{1.2}{Reaction score} & \multrow{1.3}{Reward of product} & \multrow{2.5}{NameRxn Class} \\ & & & & \\ \hline
    \multicolumn{5}{|c|}{\bf{8 Reaction Classes} (Figure \ref{fig:parallel-chemistry})} \\ \hline
\multrow{4}{NCCc1c[nH]c2ccccc12 \\ S=C(Oc1ccccn1)Oc1ccccn1} & S=C=NCCc1c[nH]c2ccccc12 & 1.000 & -- & Amino to isothiocyanato \\ & & & & \\ \hline
\multrow{4}{NCCCCCN \\ S=C=NCCc1c[nH]c2ccccc12} & NCCCCCNC(=S)NCCc1c[nH]c2ccccc12 & 0.966 & 0.841 & \multrow{4}{Isothiocyanate + \\ amine thiourea coupling} \\ & & & & \\ \hline
\multrow{4}{CCCCCCCCCCCCCCC(=O)O \\ Oc1cc(O)cc(O)c1} & CCCCCCCCCCCCCCC(=O)c1c(O)cc(O)cc1O & 0.570 & 0.827 & Friedel-Crafts acylation \\ & & & & \\ \hline
Oc1cc(Cl)c(Br)cc1Cl & Oc1c(Cl)cc(Br)c(Cl)c1Cl & 0.944 & 0.741 & Chlorination \\ \hline
OCc1ccc(O)cc1O & O=Cc1ccc(O)cc1O & 0.998 & 0.487 & Alcohol to aldehyde oxidation \\ \hline
\multrow{4}{CN(C)C1CCNC1 \\ COc1cc2nc(Cl)nc(N)c2cc1OC} & COc1cc2nc(N3CCC(N(C)C)C3)nc(N)c2cc1OC & 1.000 & 0.501 & Chloro N-arylation \\ & & & & \\ \hline
\multrow{4}{COc1ccc(C=O)c(OC)c1OC \\ NNc1nccs1} & COc1ccc(C=NNc2nccs2)c(OC)c1OC & 0.999 & 0.485 & Aldehyde reductive imination \\ & & & & \\ \hline
\multrow{4}{Nc1ccnc2cc(Cl)ccc12 \\ O=Cc1cc(O)ccc1O} & Oc1ccc(O)c(CNc2ccnc3cc(Cl)ccc23)c1 & 1.000 & 0.818 & Aldehyde reductive amination \\ & & & & \\ 
    \hline
    \multicolumn{5}{|c|}{ \bf{1 Reaction Class (Figure \ref{fig:parallel-chemistry})} } \\ \hline 
\multrow{4}{Clc1ncns1 \\ Nc1cccc(O)c1} & Oc1cccc(Nc2ncns2)c1 & 0.850 & 0.197 & Chloro N-arylation \\ & & & & \\ \hline
\multrow{4}{Clc1ccc(Br)cn1 \\ Nc1ccc(Cl)c(Cl)c1} & Clc1ccc(Nc2ccc(Br)cn2)cc1Cl & 0.996 & 0.047 & Chloro N-arylation \\ & & & & \\ \hline
\multrow{4}{Cc1ccc(N)cc1 \\ Clc1ccc2c(Cl)ccnc2c1} & Cc1ccc(Nc2ccnc3cc(Cl)ccc23)cc1 & 0.999 & 0.080 & Chloro N-arylation \\ & & & & \\ \hline
\multrow{4}{CN(C)C1CCNC1 \\ COc1cc2nc(Cl)nc(N)c2cc1OC} & COc1cc2nc(N3CCC(N(C)C)C3)nc(N)c2cc1OC & 1.000 & 0.501 & Chloro N-arylation \\ & & & & \\ \hline
\multrow{4}{Fc1ccc(OC2CCNCC2)c(F)c1 \\ Nc1ccnc(Cl)c1[N+](=O)[O-]} & Nc1ccnc(N2CCC(Oc3ccc(F)cc3F)CC2)c1[N+](=O)[O-] & 1.000 & 0.351 & Chloro N-arylation \\ & & & & \\ \hline
\multrow{4}{COc1ccc2ncc(C\#N)c(Cl)c2c1 \\ Nc1cc(Cl)ccc1Cl} & COc1ccc2ncc(C\#N)c(Nc3cc(Cl)ccc3Cl)c2c1 & 0.995 & 0.088 & Chloro N-arylation \\ & & & & \\ \hline
\multrow{4}{Cc1cc(N)cc(C)c1C \\ N\#Cc1ccc(Cl)c(C(=O)O)c1} & Cc1cc(Nc2ccc(C\#N)cc2C(=O)O)cc(C)c1C & 0.015 & 0.583 & Chloro N-arylation \\ & & & & \\ \hline
\multrow{4}{COc1ccc2c(=O)[nH]cnc2c1 \\ Fc1ccc(Cl)nc1} & COc1ccc2c(=O)n(-c3ccc(F)cn3)cnc2c1 & 0.989 & 0.056 & Chloro N-arylation \\ & & & & \\ 
     \hline
    \end{tabular}
}
\egroup
\captionof{table}{Reaction SMILES, NameRxn \cite{noauthor_hazelnut_nodate} classes, reaction likelihood scores, and reward values for the reactions shown in Figure \ref{fig:parallel-chemistry}. \label{tab:parallel_table}}

\bgroup
\def\arraystretch{0.8}
\Rotatebox{90}{
    \scriptsize
    \begin{tabular}{|l|l|l|l|l|}
    \hline
    \multrow{3}{Reactant SMILES} & \multrow{3}{Product SMILES} & \multrow{1.2}{Reaction score} & \multrow{1.3}{Reward of product} & \multrow{1.3}{Cluster of product} \\ & & & & \\ 
    \hline
    \multicolumn{5}{|c|}{\bf Parallel chemistry considerations only (Figure \ref{fig:combining})} \\ \hline
\multrow{4}{Clc1ncns1 \\ Nc1cccc(O)c1} & Oc1cccc(Nc2ncns2)c1 & 0.850 & 0.197 & 15 \\& & & & \\ \hline
\multrow{4}{Clc1ccc(Br)cn1 \\ Nc1ccc(Cl)c(Cl)c1} & Clc1ccc(Nc2ccc(Br)cn2)cc1Cl & 0.996 & 0.047 & 0 \\& & & & \\ \hline
\multrow{4}{Cc1ccc(N)cc1 \\ Clc1ccc2c(Cl)ccnc2c1} & Cc1ccc(Nc2ccnc3cc(Cl)ccc23)cc1 & 0.999 & 0.080 & 0 \\& & & & \\ \hline
\multrow{4}{CN(C)C1CCNC1 \\ COc1cc2nc(Cl)nc(N)c2cc1OC} & COc1cc2nc(N3CCC(N(C)C)C3)nc(N)c2cc1OC & 1.000 & 0.501 & 3 \\& & & & \\ \hline
\multrow{4}{Fc1ccc(OC2CCNCC2)c(F)c1 \\ Nc1ccnc(Cl)c1[N+](=O)[O-]} & Nc1ccnc(N2CCC(Oc3ccc(F)cc3F)CC2)c1[N+](=O)[O-] & 1.000 & 0.351 & 1 \\& & & & \\ \hline
\multrow{4}{COc1ccc2ncc(C\#N)c(Cl)c2c1 \\ Nc1cc(Cl)ccc1Cl} & COc1ccc2ncc(C\#N)c(Nc3cc(Cl)ccc3Cl)c2c1 & 0.995 & 0.088 & 0 \\& & & & \\ \hline
\multrow{4}{Cc1cc(N)cc(C)c1C \\ N\#Cc1ccc(Cl)c(C(=O)O)c1} & Cc1cc(Nc2ccc(C\#N)cc2C(=O)O)cc(C)c1C & 0.015 & 0.583 & 0 \\& & & & \\ \hline
\multrow{4}{COc1ccc2c(=O)[nH]cnc2c1 \\ Fc1ccc(Cl)nc1} & COc1ccc2c(=O)n(-c3ccc(F)cn3)cnc2c1 & 0.989 & 0.056 & 0 \\& & & & \\ \hline
    \multicolumn{5}{|c|}{ \bf Parallel chemistry and diversity considerations (Figure \ref{fig:combining})}  \\ \hline 
\multrow{4}{Clc1ncns1 \\ Nc1cccc(O)c1} & Oc1cccc(Nc2ncns2)c1 & 0.850 & 0.197 & 15 \\& & & & \\ \hline
\multrow{4}{CN(C)C1CCNC1 \\ COc1cc2nc(Cl)nc(N)c2cc1OC} & COc1cc2nc(N3CCC(N(C)C)C3)nc(N)c2cc1OC & 1.000 & 0.501 & 3 \\& & & & \\ \hline
\multrow{4}{Cn1nc(C(C)(C)C)cc1Cl \\ NCc1ccc([N+](=O)[O-])cc1} & Cn1nc(C(C)(C)C)cc1NCc1ccc([N+](=O)[O-])cc1 & 0.999 & 0.001 & 2 \\& & & & \\ \hline
\multrow{4}{Fc1ccc(OC2CCNCC2)c(F)c1 \\ Nc1ccnc(Cl)c1[N+](=O)[O-]} & Nc1ccnc(N2CCC(Oc3ccc(F)cc3F)CC2)c1[N+](=O)[O-] & 1.000 & 0.351 & 1 \\& & & & \\ \hline
\multrow{4}{Cc1cc(N)cc(C)c1C \\ N\#Cc1ccc(Cl)c(C(=O)O)c1} & Cc1cc(Nc2ccc(C\#N)cc2C(=O)O)cc(C)c1C & 0.015 & 0.583 & 0 \\& & & & \\ \hline
\multrow{4}{NCCC1CCCCC1 \\ O=C1NC(=O)C(Cl)=C1Cl} & O=C1NC(=O)C(NCCC2CCCCC2)=C1Cl & 0.908 & 0.029 & 20 \\& & & & \\ \hline
\multrow{4.5}{CN1CCNCC1 \\ O=[N+]([O-])c1ccc2nc(Cl)sc2c1} & CN1CCN(c2nc3ccc([N+](=O)[O-])cc3s2)CC1 & 1.000 & 0.002 & 12 \\& & & & \\ \hline
\multrow{4}{C1COCCN1 \\ Clc1ncnc(N2CCCCC2)n1} & c1nc(N2CCCCC2)nc(N2CCOCC2)n1 & 0.999 & 0.002 & 7 \\& & & & \\ \hline
    \end{tabular}
}
\egroup
\captionof{table}{Reaction SMILES, reaction likelihood scores, reward values, and cluster assignments for the selected reactions and selected candidates shown in Figure \ref{fig:combining}. All reactions are classified as chloro N-arylation reactions by NameRxn \cite{noauthor_hazelnut_nodate}. \label{tab:combining_table}}

\clearpage

\end{suppinfo}

\end{document}